\documentclass[%
reprint,
amsmath,amssymb,
superscriptaddress,
aps,
prb,
]{revtex4-2}
\usepackage{newtxtext,newtxmath} 
\usepackage{graphicx}
\usepackage{dcolumn}
\usepackage{bm}
\usepackage[colorlinks,linkcolor=blue,
anchorcolor=blue,
citecolor=blue,
urlcolor=blue,
colorlinks=true,
breaklinks=true]{hyperref}
\usepackage{physics}
\usepackage[version=4]{mhchem}
\usepackage{easyReview}
\usepackage{soul}
\usepackage{upgreek}
\usepackage{booktabs}
\usepackage{enumitem}

\emergencystretch=\maxdimen
\begin{document}

\preprint{APS/123-QED}

\title{Boundary-induced medium mapping enables air-equivalent acoustic propagation and perfect absorption in water}

\author{Mingyu Duan}
 \email{mingyuduan@outlook.com}
\affiliation{%
Department of Mechanics, Beijing University of Technology, Beijing 100124, China
}%
\author{Xiangjun Peng}
\email{xiangjunpeng@wustl.edu}
\affiliation{%
	Department of Engineering Mechanics, AML, Institute of Biomechanics and Medical Engineering, Tsinghua University, Beijing 100084, China
}

\author{Ying-Jing Qian}
\email{candiceqyj@163.com}
\affiliation{%
	Department of Mechanics, Beijing University of Technology, Beijing 100124, China
}%

\author{Tian Jian Lu}
\email{tjlu@nuaa.edu.cn}
\affiliation{State Key Laboratory of Mechanics and Control for Aerospace Structures, Nanjing University of Aeronautics and Astronautics, Nanjing, 210016, China}
\affiliation{MIIT Key Laboratory of Multifunctional Lightweight Materials and Structures, Nanjing University of Aeronautics and Astronautics, Nanjing, 210016, China}

\date{\today}

\begin{abstract}
The extreme water--air impedance contrast ($\sim$3600) has long acted as a fundamental barrier separating airborne and underwater acoustics.
Here, we overcome this barrier with flexible boundaries, which establish a direct physical mapping between disparate acoustic media, enabling a water-filled channel to emulate \textit{air-equivalent} wave propagation.
Through vibroacoustic coupling, the effective wave velocity is rescaled in an approximately nondispersive manner, producing slow waves with tunable attenuation.
Leveraging this concept, we demonstrate broadband underwater sound absorption at a deep-subwavelength thickness approaching the causal limit.
Our findings reveal that acoustic media can be reshaped via boundary dynamics rather than bulk composition, paving the way for transplanting acoustic functionalities and metamaterial design across distinct media.
\end{abstract}

\maketitle
\section{Introduction}
Acoustic wave manipulation is fundamentally constrained by the intrinsic properties of the host medium \cite{Kinsler_2000_Fundamentals}.
The extreme impedance mismatch between water and air ($\sim$3600 times) creates a pronounced medium barrier, leading to drastically different acoustic behaviors \cite{Bok_2018_PRL,Dong_2022_NSR}.
This disparity is especially evident in sound absorption.
In air, efficient low-frequency absorption arises from fluid resonances such as Helmholtz and Fabry--P\'erot modes \cite{Zhang_2016_PRAppl,Yang_2017_ARMR,Yang_2017_MH,Huang_2020_SB,Huang_2023_PRAppl}.
This capability stems from the low wave velocity, high compressibility, and strong viscous dissipation of air, enabling deep-subwavelength designs of metamaterials \cite{Ge_2025_PRL}.
Complementing these fluid-resonant strategies, recent work shifted the dominant dissipation from air motion to viscoelastic structural damping, realizing broadband low-frequency absorption with a single unit \cite{Zhang_2025_PNAS}.
In water, however, the high wave velocity, weak compressibility, and low attenuation suppress the above mechanisms relying on dissipation in the background medium and make low-frequency absorption intrinsically difficult \cite{Kim_2022_AA,Assouar_2018_NRM,Dong_2022_NSR}.
Consequently, underwater sound absorbers typically rely on solid resonances through bulk modification (e.g., viscoelastic materials with solid inclusions or voids \cite{Meng_2012_JSV,Gu_2021_PRAppl,Feng_2022_PRAppl,Qu_2022_SA,Dong_2023_PRAppl,Zheng_2025_MSSP}), leading to bulky, heavy structures, especially at low frequencies ($<1000~\mathrm{ Hz}$).
This limitation originates from the intrinsic mismatch in wave properties between water and air, rather than from specific structural designs.
Bridging this gap therefore requires establishing a physical mapping that allows wave propagation in water to emulate that in air.
Such a mapping would allow acoustic functionalities traditionally restricted to air to be physically transplanted into underwater environments, while dramatically reducing the size and operating frequency of acoustic metamaterials.

\begin{figure}[h]
	\centering
	\includegraphics[width=8.5 cm]{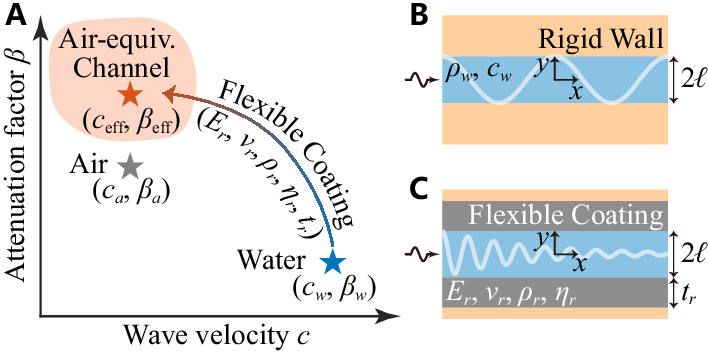}
	\caption{(A) Parameter space mapping of intrinsic wave velocity and attenuation factor $(c,\beta)$ for air/water and the water-filled \textit{air-equivalent} channel. The arrow indicates the targeted parameter shifting via flexible coatings. (B) Wave propagation in a rigid-walled channel with half-width $\ell$. (C) Wave propagation and attenuation in a channel with half-width $\ell$ and flexible coating thickness $t_r$. The white sinusoidal curves represent the spatial distribution of the acoustic pressure field within the channel.}
	\label{fig:Fig1} 
\end{figure}

In this study, we establish a boundary-induced medium mapping framework, in which the intrinsic propagation properties of an acoustic medium are deterministically transformed via tailored boundary conditions.
This paradigm shifts the design focus of metamaterials from bulk modification or pathway control \cite{Liang_2021_PRL} to boundary dynamics, enabling a deterministic rescaling of wave propagation properties fundamentally inaccessible through conventional bulk designs.
By replacing rigid walls with tailored flexible boundaries, we realize a water-filled channel exhibiting wave propagation properties equivalent to those of air via a nondispersive slow-wave effect.
The effective wave velocity $c$ and attenuation factor $\beta$ are thereby deterministically mapped through boundary parameters, establishing a direct correspondence between water and air in the $(c,\beta)$ space [Fig.\ \ref{fig:Fig1}A].
This mapping enables airborne absorption mechanisms to be directly transferred to underwater environments, yielding similar wave responses in both media using the same structure and frequency band.
To demonstrate this concept, we design an acoustic sink that achieves quasi-perfect absorption over 2.3 octaves at low frequencies, with a thickness only $7\%$ above the causality limit.

\section{Results}
\subsection{Design guideline for air-equivalent channel}
We consider a water-filled channel of half-width $\ell$ in 2D planar geometry as the model system.
As shown in Fig.\ \ref{fig:Fig1}B, a rigid-walled channel supports acoustic propagation governed by the intrinsic properties of water, where the wave velocity remains high and dissipation is weak.
In contrast, by lining the channel with a flexible rubber coating of thickness $t_r$ [Fig.\ \ref{fig:Fig1}C], the propagation characteristics can be fundamentally altered through boundary dynamics, simultaneously rescaling the effective wave velocity and attenuation.
The acoustic pressure $p$ in the flexible-walled channel satisfies the wave equation, $c_0^2 \nabla^2 p = \partial^2 p /\partial t^2$ with $c_0$ the intrinsic wave velocity of the fluid medium. 
The inviscid wave equation is utilized here since the energy dissipation is dominated by the viscoelasticity of the rubber coating, as demonstrated in the SI Appendix, Finite element model and verification.
The displacement field $\mathbf{u}$ in the rubber coating satisfies the Navier equation
$\rho_r\partial^2 \mathbf{u}/\partial t^2	=	(\lambda_r + 2\mu_r)\nabla(\nabla\cdot \mathbf{u})	-	\mu_r \nabla \times (\nabla \times \mathbf{u})$, where $\lambda_r$ and $\mu_r$ are the Lam\'e constants, and $\rho_r$ is the rubber density.
Considering the coupling between acoustic waves and coating vibrations, the acoustic pressure $p$ and $y$-direction vibration velocity $v_y$ at the flexible walls ($y=\pm \ell$) are linked by the following boundary condition:
\begin{align}\label{eq:BC}
	\left.v_y \right|_{y=\pm\ell}=\pm\frac{p}{Z_r},
\end{align}
where $Z_r=-j \rho_r c_L \cot\left(\omega t_r/c_L\right)$ is the surface impedance of the flexible rubber coating, $\omega=2\pi f$ is the angular frequency, $c_L=\sqrt{K_r\left(1+j\eta_r\right)/\rho_r}$ is the complex longitudinal wave velocity, $K_r=E_r(1-\nu_r)/[(1+\nu_r)(1-2\nu_r)]$ is the longitudinal modulus, $E_r$ is the Young's modulus, $\nu_r$ is the Poisson's ratio, and $\eta_r$ is the loss factor.
The coating dynamics is thus described by the longitudinal component alone, as the contribution of shear deformation is negligible in the long-wavelength and thin-layer limit, which is confirmed by the numerical validation and the asymptotic analysis in Appendix \ref{sec:methods}.

To characterize the relevant length scales in the vibroacoustic coupling system, we introduce three wavenumbers: the longitudinal wavenumber in the rubber coating $k_L = \omega/c_L$, the intrinsic wavenumber in water $k_0 = \omega/c_0$, and the $y$-direction wavenumber in water $k_y$.
Within the long-wavelength and thin-layer limit, $k_y \ell\ll 1$ and $k_L t_r\ll 1$ (see the derivation in Appendix \ref{sec:methods}), we obtain the $x$-direction effective phase velocity of the primary wave in the channel as
\begin{align}\label{eq:c_eff}
	\frac{c_{\mathrm{eff}}}{c_0}\approx\left(1 + \frac{ t_r}{\ell} \frac{\rho_0 c_0^2}{\rho_r c_L^2}\right)^{-\frac{1}{2}},
\end{align}
where $\rho_0$ is the density of water.
Eq.\ \eqref{eq:c_eff} is frequency-independent when the boundary material properties are treated as nondispersive.
It establishes a direct link between $c_{\mathrm{eff}}$ and the boundary parameters $t_r$, $\rho_r$, $c_L(E_r, \nu_r, \eta_r)$ and channel half-width $\ell$.
The attenuation factor $\beta_{\mathrm{eff}}$ arises from the imaginary part in $c_L$, allowing dissipation to be tuned by $\eta_r$.
Notably, such decoupled control of $c_{\mathrm{eff}}$ and $\beta_{\mathrm{eff}}$ is fundamentally inaccessible through conventional bulk material modification, where these properties are intrinsically coupled.

Interestingly, Eq.\ \eqref{eq:c_eff} bears a formal resemblance to the classical Moens--Korteweg (MK) equation for wave propagation in fluid-filled elastic tubes \cite{Korteweg_1878_AdP}, which has been widely used in arterial hemodynamics to relate pulse-wave velocity to vascular elasticity \cite{Ma_2018_PNAS}.
This similarity arises from a shared compliance-induced mechanism, although the underlying boundary dynamics are distinct.
While MK waves are governed by circumferential dilation of compliant tube walls, our flexible coating undergoes thickness-direction compression, such that increasing $t_r$ enhances compliance and reduces $c_{\mathrm{eff}}$, opposite to the MK trend.
Beyond this formal analogy, Eq.\ \eqref{eq:c_eff} establishes boundary compliance and resistance as deterministic design parameters for medium mapping, enabling the propagation characteristics of water to be engineered toward those of a target acoustic medium.

Accordingly, the acoustic pressure varies as
\begin{align}\label{eq:p_p_0}
	\frac{p}{p_0}\approx\exp\left[-j\left(1 + \frac{ t_r}{\ell} \frac{\rho_0 c_0^2}{\rho_rc_L^2}\right)^{\frac{1}{2}} k_0 x\right],
\end{align} 
where $p_0$ is the acoustic pressure of the excitation, and the imaginary part of $c_L$ accounts for the acoustic energy dissipation in the flexible wall.

\begin{figure}[h]
	\centering
	\includegraphics[width=8.5 cm]{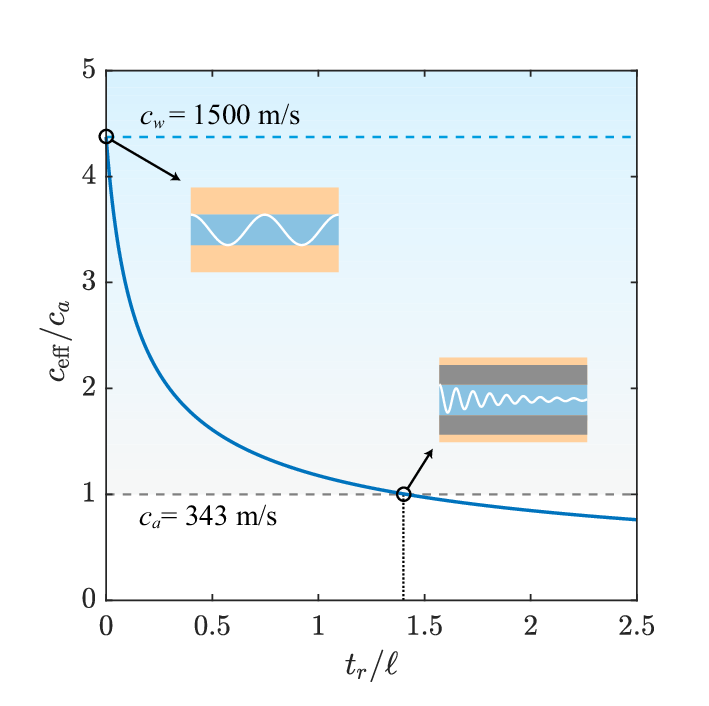}
	\caption{Dimensionless effective wave velocity $c_{\mathrm{eff}}/c_a$ as a function of the dimensionless coating thickness $t_r/\ell$, predicted by Eq.\ \eqref{eq:c_eff}. The horizontal dashed lines denote the intrinsic wave velocities in air ($c_a=343~\mathrm{m/s}$) and water ($c_w=1500~\mathrm{m/s}$). Unless otherwise specified, the material parameters are $\rho_0=1000~\mathrm{kg/m^3}$, $c_0=1500~\mathrm{m/s}$, $E_r=10~\mathrm{MPa}$, $\nu_r=0.49$, $\rho_r=1100~\mathrm{kg/m^3}$, $\eta_r=0.1$.}
	\label{fig:Fig2_1} 
\end{figure}

Eq.\ \eqref{eq:c_eff} provides a quantitative design guideline for the flexible coating thickness to match the acoustic wave velocity in the water-filled channel ($c_w= 1500~\mathrm{m/s}$) to that in air ($c_a= 343~\mathrm{m/s}$).
As shown in Fig.\ \ref{fig:Fig2_1}, the dimensionless effective wave velocity $c_{\mathrm{eff}}/c_a$ is plotted against the dimensionless coating thickness $t_r/\ell$ using typical parameters for silicone rubber \cite{Yang_2022_JSV,Liang_2025_IJMS}.
As the wall thickness increases, the effective velocity drops from $c_w$ to values below $c_a$.
An \textit{air-equivalent} condition is achieved at $t_r/\ell = 1.4$, where $c_{\mathrm{eff}} = c_a$.
Under this scenario, the channel operates as an \textit{air-equivalent} channel, with its acoustic wave propagation properties becoming analogous to those of an air-filled rigid channel.
Here, ``rigid" denotes acoustic rigidity resulting from the severe impedance mismatch between air and the rubber walls.

\subsection{Vibroacoustic coupling mechanism}
\begin{figure}[h]
	\centering
	\includegraphics[width=8.5 cm]{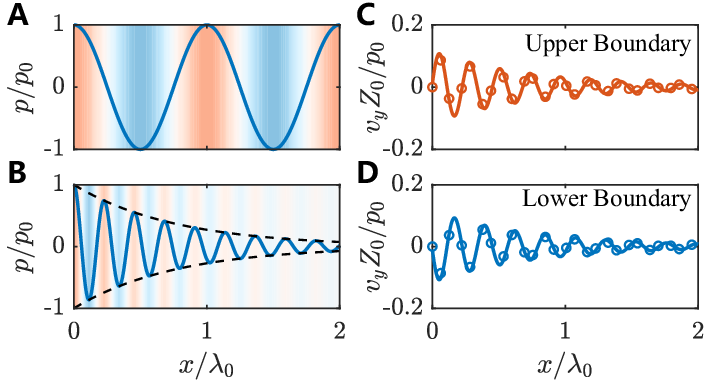}
	\caption{(A, B) Distributions of dimensionless acoustic pressure $p/p_0$ for water-filled rigid channel and \textit{air-equivalent} channel, where the dashed lines denote the attenuation of the acoustic wave amplitude. (C, D) Dimensionless vibration velocity along the $y$-direction $v_y Z_0/p_0$ at the upper and lower boundaries. Lines and colormaps/symbols ($\circ$) denote analytical and numerical predictions, respectively. Parameters used here are identical to those in Fig.\ \ref{fig:Fig2_1}.}
	\label{fig:Fig2_2} 
\end{figure}

The inset schematics in Fig.\ \ref{fig:Fig2_1} illustrate the acoustic pressure distributions in the channel, with quantitative details presented in Figs.\ \ref{fig:Fig2_2}A and B. Our theoretical predictions obtained using Eq.\ \eqref{eq:p_p_0} are in excellent agreement with numerical simulations performed in COMSOL Multiphysics for a channel with $\ell=2~\mathrm{mm}$. As shown in Fig.\ \ref{fig:Fig2_2}A, the wavelength in the rigid-walled channel is equal to the intrinsic wavelength in water $\lambda_0$, with no amplitude attenuation. In contrast, the \textit{air-equivalent} channel exhibits a typical slow-wave phenomenon. The effective wavelength is reduced to approximately $\lambda_0/4.37$, corresponding to a phase velocity of $c_a=343~\mathrm{m/s}$. In addition, the acoustic pressure amplitude $p_A$ decays monotonically along channel length. This decay is well-described by the black dashed lines, which follow the relation $p_A/p_0 = \exp[\Im(k_{\mathrm{eff}})x]$, where $k_{\mathrm{eff}}=\omega/c_{\mathrm{eff}}$ is the effective wavenumber.

Fundamentally, the slow-wave phenomenon in the \textit{air-equivalent} channel originates from the vibroacoustic coupling between the water and the flexible walls.
Substituting Eq.\ \eqref{eq:p_p_0} into Eq.\ \eqref{eq:BC}, we obtain the analytical predictions for the dimensionless vibration velocity $v_y Z_0/ p_0$ at the upper ($y=\ell$) and lower ($y=-\ell$) boundaries, as plotted by the solid lines in Figs.\ \ref{fig:Fig2_2}C and D.
Here, $Z_0=\rho_0 c_0$ is the characteristic impedance of the fluid.
These results show excellent agreement with fully coupled acoustic--structure finite element (FE) simulation, represented by symbols.
The particle velocity at the water--flexible wall interface exhibits a decaying sinusoidal form along the $x$-direction.
A $\pi/2$ phase shift exists between the vibration velocity and the acoustic pressure (cosine wave), since the wall reactance dominates the surface acoustic impedance.
The wall vibration modifies wave propagation through two mechanisms. First, the wall reactance increases the effective compliance for fluid compression and expansion in the channel, shortening the wavelength and rescaling the phase velocity. Second, the viscoelastic resistance converts acoustic energy into heat, leading to wave attenuation.
Consequently, in Fig.\ \ref{fig:Fig2_2}B, the wavelength approaches that in air, accompanied by simultaneous attenuation.
Ultimately, we establish a mapping from a water-filled to an \textit{air-equivalent} channel via flexible walls, as shown in Fig.\ \ref{fig:Fig1}A, where the effective wave velocity $c_\mathrm{eff}$ is governed by the wall modulus $E_r$, Poisson's ratio $\nu_r$, density $\rho_r$ and thickness $t_r$, and the attenuation factor $\beta_\mathrm{eff}$ can be tuned predominantly via the loss factor $\eta_r$.
Importantly, this mechanism originates from boundary dynamics rather than geometric confinement, and is therefore independent of specific structural configurations, allowing it to be generalized, in principle, to other flexible-boundary waveguiding systems.

\begin{figure*}[t]
	\centering
	\includegraphics[width=12.5 cm]{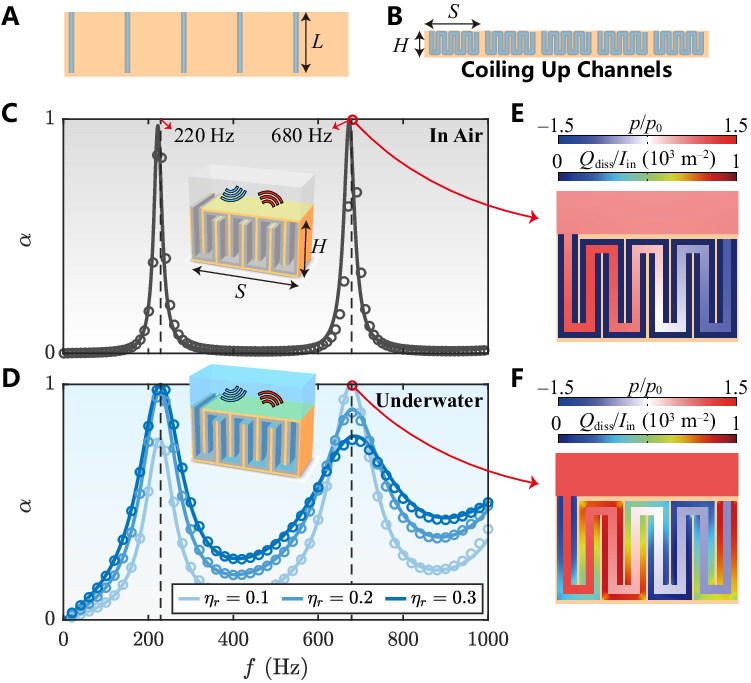}
	\caption{(A) Schematic of straight \textit{air-equivalent} channels with length $L=377~\mathrm{mm}$. (B) Coiling up channels with cell height $H=50~\mathrm{mm}$, cell length $S=84.8~\mathrm{mm}$ and number of folds $n=8$. (C, D) Sound absorption coefficient $\alpha$ of the same coiling up channel in air and water for different $\eta_r$. Lines and symbols ($\circ$) denote analytical and numerical predictions, respectively. (E, F) Distributions of dimensionless acoustic pressure $p/p_0$ and energy dissipation density $Q_{\mathrm{diss}}/I_{\mathrm{in}}$ at the second resonance frequency, where $I_{\mathrm{in}}$ is the incident acoustic intensity. Parameters used here are identical to those in Fig.\ \ref{fig:Fig2_1}, except for $\eta_r$.}
	\label{fig:Fig3} 
\end{figure*}

\section{Discussion}
\subsection{Sound absorption in dual-medium}
Using the concept of coiling up space \cite{Liang_2021_PRL}, we illustrate this mapping through a deep-subwavelength resonator that achieves perfect sound absorption in both air and water within the same frequency band [Figs.\ \ref{fig:Fig3}A and B].
The surface acoustic impedance of the coiling up channel, with number of folds $n$, channel length $L$ and cell length $S$, is given by $Z_s=-j \rho_0 c_{\mathrm{eff}} \cot(k_{\mathrm{eff}}L)S/(2\ell)$. The sound absorption coefficient with rigid backing is then determined by $\alpha = 1 - |(Z_s - Z_0)/(Z_s + Z_0)|^2$.
The air-filled rigid channel in Fig.\ \ref{fig:Fig3}C is modeled using the theory of microslit absorbers \cite{Maa_2000_AA}, incorporating viscous dissipation in air (see SI Appendix, Theory of microslit absorbers).
FE simulations using $\ell=2~\mathrm{mm}$ corroborate the analytical predictions, as shown in Figs.\ \ref{fig:Fig3}C and D.
Full 3D FE simulations confirm that the analytical model based on 2D geometry accurately captures the essential physics, including the \textit{air-equivalent} slow-wave effect and the perfect absorption (see SI Appendix, Finite element model and verification).
The coiling up channel functions as a Fabry--P\'erot resonator, with multiband resonances at frequencies corresponding to odd multiples of $1/4$ wavelength.
Accordingly, the phase shift from the channel entrance to the end satisfies $\mathrm{Re}(k_{\mathrm{eff}}L)\approx(2q-1)\pi/2$, producing quarter-wave standing-wave patterns, as shown in Figs.\ \ref{fig:Fig3}E and F.
At approximately $220~\mathrm{Hz}$ and $680~\mathrm{Hz}$, corresponding to $L\approx\lambda_{\mathrm{eff}}/4$ and $L\approx3\lambda_{\mathrm{eff}}/4$, respectively, the coiling up channel exhibits perfect absorption in both air and water.
Notably, the thickness of the coiling up channel is only $H = 50~\mathrm{mm}$, which is $1/136$ of the wavelength in water at $220~\mathrm{Hz}$.
This deep-subwavelength nature originates from three factors: \textit{i)} a slow-wave effect shortening the effective wavelength by $\lambda_{\mathrm{eff}}=\lambda_w/4.37$, \textit{ii)} a Fabry--P\'erot resonance occurring at $L=\lambda_{\mathrm{eff}}/4$, and \textit{iii)} a space-coiling design reducing the thickness to $H\approx L/8$.

A counterintuitive observation is that the absorption bandwidth of the channel is broader when it is immersed in water than when filled with air.
This difference results from acoustic impedance matching.
In air, the severe impedance mismatch with rubber renders the flexible wall effectively rigid, thereby suppressing wall vibrations.
As a result, energy dissipation from the wall is negligible, as shown in Fig.\ \ref{fig:Fig3}E.
The majority of the absorption arises from viscous shear at the wall surface, a mechanism that becomes minimal when the channel half-width $\ell$ significantly exceeds the viscous boundary layer thickness.
In contrast, when the channel is immersed in water, the comparable impedance of rubber and water permits substantial wall vibration in response to acoustic excitation, as discussed in Figs.\ \ref{fig:Fig2_2}C and D.
Vibroacoustic coupling then transfers energy from the longitudinal wave in water into wall compression and expansion, thereby enhancing energy dissipation, as shown in Fig.\ \ref{fig:Fig3}F.
Our results further demonstrate that the absorption bandwidth and damping regime can be tuned by varying the loss factor $\eta_r$ of the rubber, controlled via material composition and microstructure \cite{Wang_2020_JPCL}.

\begin{figure*}[t]
	\centering
	\includegraphics[width=14 cm]{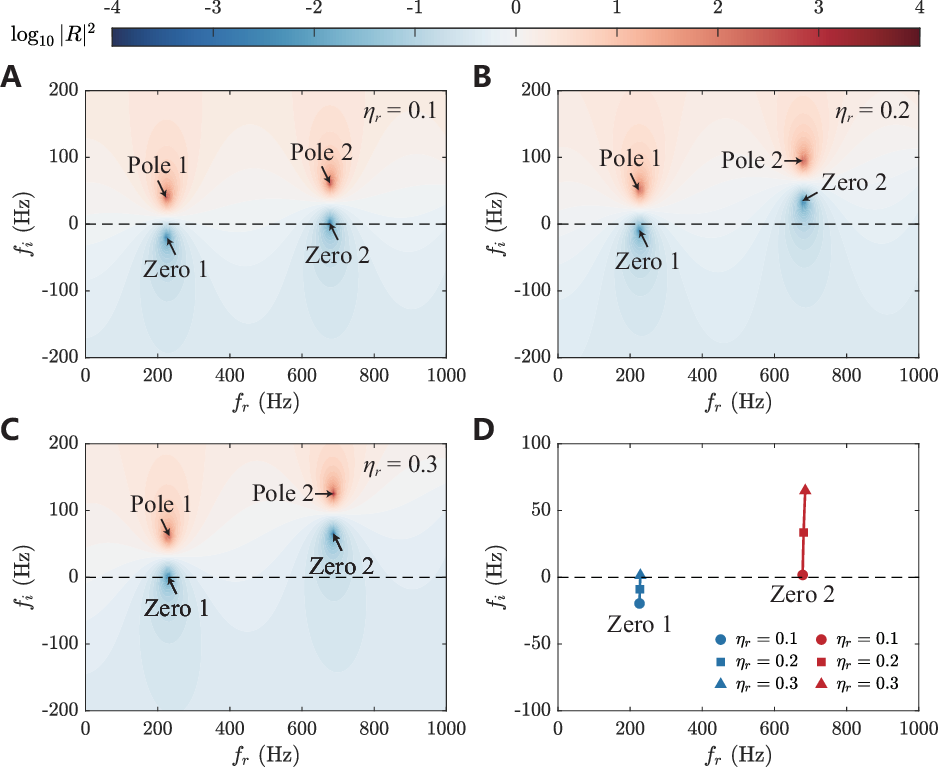}
	{\caption{(A, B, C) Representation of the $\log_{10}|R|^2$ in the complex frequency plane for the coiling up channel shown in Fig.\ \ref{fig:Fig3}B for $\eta_r=0.1$, $\eta_r=0.2$ and $\eta_r=0.3$. The arrows mark the zeros (local minima) and poles (local maxima) of $\log_{10}|R|^2$.
			(D) Trajectories of zeros in the complex frequency plane with different loss factor $\eta_r$. Markers represent the zeros for $\eta_r=0.1$, $\eta_r=0.2$ and $\eta_r=0.3$, respectively. Parameters used here are identical to those in Fig.\ \ref{fig:Fig3}.}
		\label{fig:complex_frequency_plane} }
\end{figure*}

\subsection{Complex frequency plane analysis}
{To quantitatively explain the underwater absorption mechanism and identify the damping regimes of the Fabry--P\'erot resonances in Fig.\ \ref{fig:Fig3}D, we perform a complex frequency plane analysis of the reflection coefficient $R$.
	We extend the frequency into the complex domain as $f=f_r+j f_i$ and map $\log_{10}|R|^2$ in the complex frequency plane, as illustrated in Figs.\ \ref{fig:complex_frequency_plane}A--C.
	In this framework, the zeros and poles represent absolute minima and maxima of the reflection coefficient.
	The imaginary part of these points ($f_i$) provides a fundamental quantitative measure of the balance between external radiation leakage and internal dissipation.}

The damping regime of a resonance is deterministically governed by the vertical position of the zeros relative to the real frequency axis ($f_i=0$) \cite{Romero_2016_JASA}:
\begin{itemize}[noitemsep,topsep=0pt]
	\item Underdamped: When a zero lies below the real axis ($f_i < 0$), the internal loss is insufficient to compensate for radiation leakage.
	\item Critically damped: When a zero lies exactly on the real axis ($f_i = 0$), the internal loss and radiation leakage are perfectly balanced, yielding perfect absorption ($\alpha = 1$).
	\item Overdamped: When a zero lies above the real axis ($f_i > 0$), the internal losses exceed the level required for radiation leakage. Although the absorption bandwidth is broadened, the peak value is reduced.
\end{itemize}
Figures \ref{fig:complex_frequency_plane}A--C illustrate the evolution of the zero-pole pairs associated with the first two Fabry--P\'erot resonance modes as the rubber loss factor $\eta_r$ varies from 0.1 to 0.3, with the corresponding damping regimes summarized in Table \ref{tab:damping_states}.
The trajectory of Zero 1 (approximately 220 Hz) is traced by the blue line and markers in Fig.\ \ref{fig:complex_frequency_plane}D.
At $\eta_r=0.1$ and 0.2, Zero 1 is located below the real axis ($f_i < 0$), indicating an underdamped regime where $\alpha<1$.
As $\eta_r$ increases to 0.3, Zero 1 migrates vertically and intersects the real axis ($f_i = 0$), signifying that the mode has reached critical damping and perfect absorption.
This vertical migration quantitatively explains the enhancement of the first absorption peak in Fig.\ \ref{fig:Fig3}D with the increase of $\eta_r$.

\begin{table}[h]
	\centering
	\caption{Damping regimes for the first two Fabry--P\'erot resonance modes (Zero 1 and Zero 2) with varying rubber loss factors $\eta_r$.}
	\label{tab:damping_states}
	\begin{ruledtabular}
	\begin{tabular}{ccc}
		& Zero 1 ($\sim$220 Hz) & Zero 2 ($\sim$680 Hz) \\
		\midrule
		$\eta_r = 0.1$ & Underdamped & Critically damped \\
		$\eta_r = 0.2$ & Underdamped & Overdamped \\
		$\eta_r = 0.3$ & Critically damped & Overdamped \\
	\end{tabular}
	\end{ruledtabular}
\end{table}

\begin{figure*}[t]
	\centering
	\includegraphics[width=16 cm]{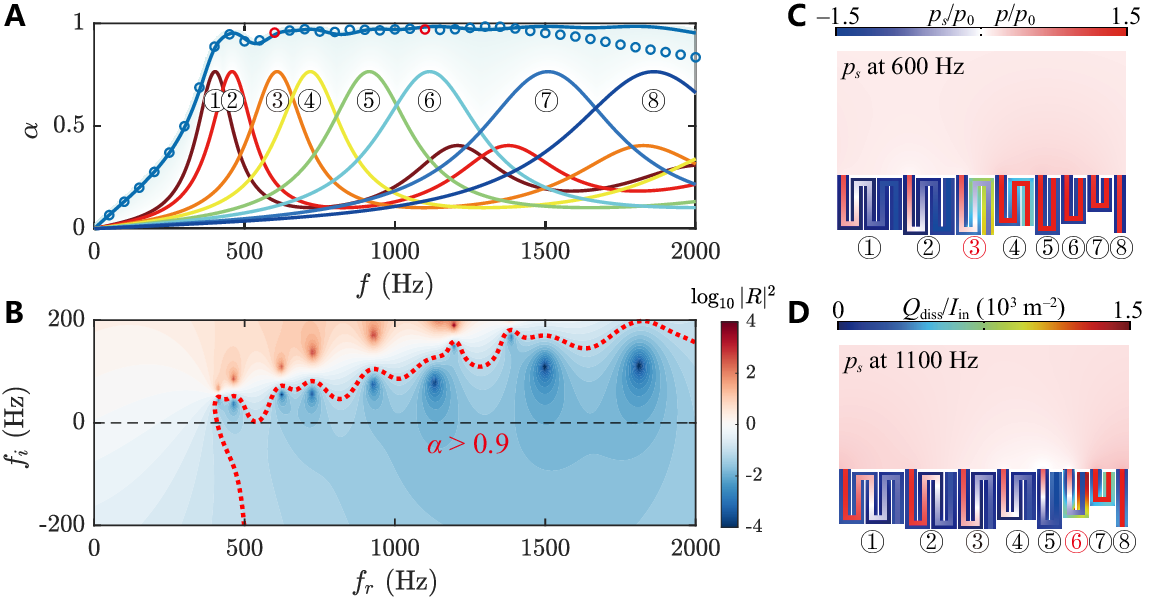}
	\caption{(A) Underwater sound absorption coefficient $\alpha$ for a metasurface composed of eight coiling up channels. Lines and symbols ($\circ$) denote analytical and numerical predictions, respectively.
		(B) Representation of the $\log_{10}|R|^2$ in the complex frequency plane for the metasurface. The red dashed curve denotes the $\alpha = 0.9$ contour.
		(C, D) Distributions of dimensionless scattered acoustic pressure $p_s/p_0$, acoustic pressure in the channels $p/p_0$, and normalized energy dissipation density $Q_{\mathrm{diss}}/I_{\mathrm{in}}$ in the coating at $600~\mathrm{Hz}$ and $1100~\mathrm{Hz}$, respectively. Parameters used here are identical to those in Fig.\ \ref{fig:Fig2_1}, except for $\eta_r=0.3$.}
	\label{fig:Fig4} 
\end{figure*}

The trajectory of Zero 2 (approximately 680 Hz) is plotted by the red line and markers in Fig.\ \ref{fig:complex_frequency_plane}D.
At $\eta_r=0.1$, Zero 2 is located exactly on the real axis ($f_i = 0$), achieving critical damping and perfect absorption.
However, as $\eta_r$ further increases to 0.2 and 0.3, Zero 2 crosses the real axis and moves into the overdamped region ($f_i>0$).
Under this scenario, the internal losses exceed the radiation leakage, causing the peak absorption value to drop below 1.

In summary, the complex frequency plane analysis confirms that the rubber loss factor $\eta_r$ serves as a deterministic regulator of the damping regime.
By tailoring $\eta_r$, a transition from an underdamped to a critically damped regime can be achieved, thereby enabling perfect absorption through the precise balance of leakage and loss.

\begin{figure*}[t]
	\centering
	\includegraphics[width=11.4 cm]{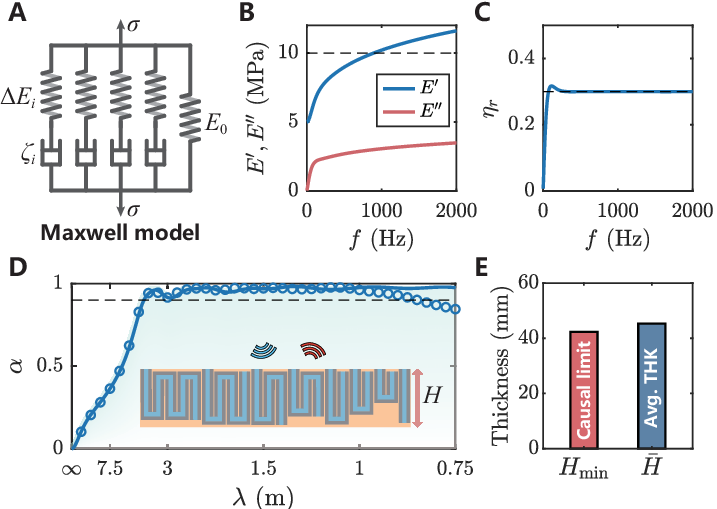}
	\caption{(A) Causal generalized Maxwell model for rubber, consisting of an equilibrium spring $E_0$ connected in parallel with four Maxwell branches. Each branch contains a spring of modulus $\Delta E_i$ and a dashpot of viscosity $\zeta_i$ connected in series.
		(B) Frequency-dependent storage modulus $E^\prime$ and loss modulus $E^{\prime\prime}$ predicted by the Maxwell model. 
		(C) Frequency-dependent loss factor $\eta_r=E^{\prime\prime}/E^{\prime}$.
		The dashed lines in (B) and (C) denote the reference values $E'=10~\mathrm{MPa}$ and $\eta_r=0.3$, respectively.
		(D) Underwater sound absorption spectrum $\alpha(\lambda)$, where the line and circles denote the analytical and numerical predictions, respectively.
		The horizontal dashed line indicates $\alpha=0.9$.
		The inset illustrates the metasurface composed of eight coiling up channels with thickness $H$.
		(E) Comparison between the causality-constrained minimum thickness $H_{\min}$ and the average physical thickness $\bar{H}$.}
	\label{fig:Maxwell_model} 
\end{figure*}

\subsection{Broadband underwater sound absorption}
As a demonstration of the generality of the acoustic parameter mapping, we consider a subwavelength acoustic sink that realizes broadband low-frequency absorption in water.
The metasurface, comprising eight parallel coiling up \textit{air-equivalent} channels (parameters in Table \ref{tab:parameters}), functions as an effective broadband impedance-matched boundary.
The structural parameters are obtained from a genetic algorithm (GA) to maximize the average absorption coefficient over 400--2000 Hz (see SI Appendix, Optimization setup for the genetic algorithm).
Note that GA serves merely as an optimization tool to identify the optimal parameter sets, whereas the underlying boundary-induced medium mapping framework is independent of the particular optimization algorithm and specific structural designs.

\begin{table}[h]
	\caption{\label{tab:parameters}%
		Structural parameters for $i$-th \textit{air-equivalent} channel.\footnote{Other parameters are $\ell=2~\mathrm{mm}$, $t_r=2.8~\mathrm{mm}$, $S=233.2~\mathrm{mm}$.}
	}
	\begin{ruledtabular}
		\begin{tabular}{l*{8}{c}}
			$i$ & 1 & 2 & 3 & 4 & 5 & 6 & 7 & 8 \\
			\midrule
			$n_i$ & 5 & 4 & 3 & 3 & 2 & 2 & 2 & 1 \\
			$L_i~\mathrm{(mm)}$ & 214.3 & 187.9 & 141.6 & 119.9 & 94.2 & 77.2 & 57.1 & 46.3 \\
			$H_i~\mathrm{(mm)}$ & 45.7 & 49.7 & 49.9 & 42.6 & 49.6 & 41.1 & 31.0 & 48.3 \\
		\end{tabular}
	\end{ruledtabular}
\end{table}

For $m$ parallel channels with individual surface impedance $Z_i$ ($i=1,2,...,m$), the total surface impedance is $Z=1/\sum_{i=1}^{m}(Z_i^{-1})$.
The analytical model predicts a quasi-perfect absorption band ($\alpha>0.9$) from 409 Hz to at least 2000 Hz, spanning approximately 2.3 octaves, as shown in Fig.\ \ref{fig:Fig4}A.
Both analytical predictions and fully coupled acoustic--structure FE simulations confirm the broadband performance.
The slight deviation between analytical and numerical results above 1500 Hz results from the increasing contribution of transverse modes beyond the long-wavelength and thin-layer approximations.

Importantly, the colored curves in Fig.~\ref{fig:Fig4}A show that none of the isolated channels reaches perfect absorption.
Complex frequency plane analysis in Fig.~\ref{fig:Fig4}B confirms that the multiple reflection zeros of the metasurface lie above the real frequency axis, indicating that its collective resonances operate in the overdamped regime.
The broadband response therefore does not result from simply combining critically coupled absorption peaks.
Instead, these imperfect, overdamped Fabry--P\'erot resonances overlap spectrally and combine coherently through the shared exterior acoustic field, thereby producing a collective, coherent coupling response \cite{Huang_2020_SB,Zhou_2022_NSR}.
Correspondingly, the overlapping low-reflection regions in the complex frequency plane merge into a continuous $\alpha>0.9$ domain intersecting the real frequency axis, as shown in Fig.~\ref{fig:Fig4}B.

The weak scattered pressure fields $p_s/p_0\approx 0$ in Figs.\ \ref{fig:Fig4}C and D confirm the nearly anechoic response of the acoustic sink.
Channel \textcircled{\small{3}} provides the dominant dissipation at 600 Hz, whereas channel \textcircled{\small{6}} dominates at 1100 Hz.
Spectrally adjacent channels nevertheless remain active and contribute to the collective surface impedance.
In contrast to high-reactance, narrowband quasi-Helmholtz resonators \cite{Duan_2021_APL,Duan_2023_AA,Zhou_2022_PLA}, the \textit{air-equivalent} channel supports broad, spectrally overlapping Fabry--P\'erot resonances, enabling broadband perfect absorption in a more compact array.

\subsection{Causality constraint on the absorber thickness}
We finally assess the thickness efficiency of the broadband absorber against the causal lower bound.
For a passive, linear, time-invariant one-port absorber under normal incidence, causality imposes the following lower bound on its thickness \cite{Yang_2017_ARMR,Yang_2017_MH} 
\begin{align}
	H\geq H_{\mathrm{min}} =\frac{1}{4\pi^2} \frac{B_\mathrm{eff}}{B_0} \left| \int_0 ^\infty \ln\left[1-\alpha \left(\lambda\right)\right]d\lambda\right|,
	\label{eq:causal_bound}
\end{align}
where $B_0=\rho_0 c_0^2$ is the bulk modulus of water, $B_{\mathrm{eff}}$ is the effective bulk modulus of the absorber at the static limit ($f\rightarrow0$), and $\lambda=c_0/f$ is the wavelength in water.

Evaluating the causal thickness bound requires a causal material constitutive model \cite{Qu_2022_SA}.
However, when extrapolated over the full frequency range, the rubber model employed above, $E^{*}=E(1+\mathrm{j}\eta_r)$, predicts a finite loss modulus even at zero frequency while keeping the storage modulus constant, thereby violating the Kramers--Kronig relations.
We therefore replace it with a passive and causal four-branch generalized Maxwell model \cite{Salenccon_2019_viscoelastic} illustrated in Fig.\ \ref{fig:Maxwell_model}A:
\begin{align}
	E^{*}\left(\omega\right)=	
	E_{0}
	+
	\sum_{i=1}^{4}
	\Delta E_i
	\frac{j\omega\tau_i}
	{1+j\omega\tau_i}.
	\label{eq:generalized_Maxwell}
\end{align}
Here, $E_0$ is the relaxed equilibrium modulus, while $\Delta E_i$ and $\tau_i=\zeta_i/\Delta E_i$ are the spring modulus and relaxation time of the $i$-th Maxwell branch, respectively, with $\zeta_i$ denoting the dashpot viscosity.
Then the longitudinal modulus and longitudinal wave velocity are $K_r^{*}(\omega)=E^{*}(\omega)(1-\nu_r)/[(1+\nu_r)(1-2\nu_r)]$ and $c_L(\omega)=\sqrt{K_r^{*}(\omega)/\rho_r}$, respectively.
The loss factor is defined as $\eta_r(\omega)=E''/E'$, where $E'$ and $E''$ denote the storage and loss moduli, respectively.
The parameters listed in Table \ref{tab:Maxwell_parameters} were fitted to reproduce the baseline $E_\mathrm{base}^*=10~\mathrm{MPa}(1+0.3j)$ over 400--2000 Hz.
As shown in Figs.\ \ref{fig:Maxwell_model}B and C, the resulting storage and loss moduli exhibit frequency dependence required by causality, while the loss factor remains approximately $0.3$ throughout the target band.

\begin{table}[h]
	\centering
	\caption{Parameters of the four-branch generalized Maxwell model.\footnote{The equilibrium modulus is $E_0=5~\mathrm{MPa}$.}}
	\label{tab:Maxwell_parameters}
	\begin{ruledtabular}
		\begin{tabular}{ccc}
			Branch & $\Delta E_i$ (MPa) & $\tau_i$ (s) \\
			\hline
			1 & $2.85393$ & $1.56147\times10^{-3}$ \\
			2 & $1.95464$ & $3.20149\times10^{-4}$ \\
			3 & $2.48241$ & $1.01371\times10^{-4}$ \\
			4 & $7.67881$ & $1.82867\times10^{-5}$ \\
		\end{tabular}
	\end{ruledtabular}
\end{table}

Using the fitted Maxwell model, the absorption spectrum $\alpha(\lambda)$ is recalculated without changing the optimized geometry, as shown in Fig.~\ref{fig:Maxwell_model}D.
The causal model closely reproduces the spectrum predicted in Fig.\ \ref{fig:Fig4}A, confirming that the latter remains an accurate in-band approximation and that the broadband absorption is preserved after enforcing a causal constitutive response.

The static modulus entering Eq.\ \eqref{eq:causal_bound} is evaluated using Wood's formula. Neglecting the compliance of the rigid supporting phase, it is given by
\begin{align}\label{eq:static_Beff}
	B_\mathrm{eff}=\left(\frac{\varphi_0}{B_0}+\frac{\varphi_r}{B_r}\right)^{-1},
\end{align}
with $\varphi_0=\ell/(\ell+t_r+t_s/2)$ and $\varphi_r=t_r/(\ell+t_r+t_s/2)$ being the volume fractions of water and rubber, respectively, accounting for the rigid partition thickness $t_s=1~\mathrm{mm}$.
Here, $B_{r}=E_0(1-\nu_r)/[(1+\nu_r)(1-2\nu_r)]$ is the relaxed longitudinal modulus of the rubber.

Substituting $B_{\mathrm{eff}}$ and the analytically calculated absorption spectrum
$\alpha(\lambda)$ into Eq.\ \eqref{eq:causal_bound} yields a causal lower bound of $H_{\min}=42.33~\mathrm{mm}$.
According to Ref.\ \cite{Qu_2022_SA}, the average thickness of the eight coiled channels is $\bar{H}=\sum_{i=1}^8 n_i H_i/\sum_{i=1}^8 n_i=45.3~\mathrm{mm}$.
Consequently, $H_{\min}/\bar H=0.935$, and the realized average thickness (Avg. THK) exceeds the causal lower bound by only $7.0\%$, as shown in Fig.~\ref{fig:Maxwell_model}E.
This small difference arises in part from the finite thickness of the metal face sheets required for structural support.

Therefore, our design achieves low-frequency, broadband underwater absorption with an average thickness close to the causal lower bound.
Pioneering studies have approached causal bounds for electromagnetic \cite{Qu_2021_PNAS}, airborne acoustic \cite{Yang_2017_ARMR,Yang_2017_MH}, and underwater acoustic absorbers \cite{Qu_2022_SA}.
Our results establish a distinct route based on an explicitly causal viscoelastic constitutive model and boundary-induced fluid--solid vibroacoustic coupling, thereby extending causal-limit design to a new class of underwater sound absorbers.

\subsection{Feasibility and robustness}
From a practical perspective, the proposed mechanism can be realized using commonly available soft materials and standard fabrication techniques.
For instance, silicone rubbers with elastic moduli of several MPa, density on the order of $10^3~\mathrm{kg/m^3}$, and tunable loss factors have been widely used in underwater acoustic coatings, providing a feasible parameter space for achieving the required boundary impedance \cite{Ba_2017_SR,Wang_2020_JPCL}.
As a representative implementation, the flexible coating can be fabricated by casting commercially available silicone rubber and subsequently integrated with a rigid supporting structure using silicone rubber adhesive \cite{Liu_2026_ACHM}.
In contrast to conventional underwater sound absorbers relying on enclosed air cavities, the present structure eliminates air inclusions and operates entirely in a fluid-filled configuration, thereby avoiding the severe volumetric collapse associated with enclosed air cavities.
The influence of hydrostatic pressure in our system primarily modifies the dynamic mechanical properties and thickness of the coating \cite{Yang_2022_JSV,Zhang_2025_PS}, which can be directly incorporated into the present framework through updated boundary parameters without altering the underlying mechanism.
In addition, the rigid-wall assumption adopted in the model can be effectively realized in practice by reinforcing the supporting structures (see SI Appendix, Finite element model and verification), which provide high stiffness-to-weight ratios and structural stability under external loading \cite{Jiang_2024_MTC}.
A supplemental robustness analysis in the SI Appendix, Robustness analysis for the broadband absorber demonstrates that the broadband absorption performance remains stable against fabrication tolerances ($\pm 10\%$) and imperfect interfacial bonding.
Although frequency- and pressure-dependent rubber properties moderately reduce the average absorption level under hydrostatic pressures up to 4 MPa, such degradation could be mitigated through reoptimization using the pressure-dependent constitutive parameters.
These results suggest that the proposed boundary-induced medium mapping is compatible with realistic material systems and underwater operating conditions.

\section{Conclusion and Outlook}

In summary, we establish a boundary-induced mapping framework that creates a deterministic correspondence between acoustic propagation behaviors in distinct media.
By introducing flexible walls, a water-filled channel is transformed into an \textit{air-equivalent} regime with simultaneously rescaled wave velocity and enhanced dissipation.
This vibroacoustic coupling produces approximately nondispersive slow-wave propagation and tunable attenuation over the target band, enabling acoustic functionalities previously restricted to airborne systems to be realized underwater.
Importantly, our work introduces a strategy that reconstructs the effective wave propagation medium through boundary dynamics, rather than relying on geometric manipulation of propagation paths.
Leveraging this concept, we demonstrate a deep-subwavelength acoustic sink that achieves low-frequency, broadband underwater sound absorption while approaching the causal thickness limit.
By shifting wave manipulation from spatial structuring to boundary-induced medium mapping, this approach opens a new paradigm for developing compact acoustic metamaterials, with potential extensions to elastic, air-porous, and other hybrid wave systems.

\appendix


	\section{Derivation of effective wave velocity in the flexible-walled channel}
	\label{sec:methods}
	We consider a water-filled channel of width $2\ell$ lined with a flexible rubber coating of thickness $t_r$, as illustrated in Fig.\ \ref{fig:Fig1}C. The effective wave velocity is obtained by solving the governing equations in the fluid and the elastic coating, subject to continuity of stress and velocity at the fluid--solid interface.
	
	\textit{i) Governing equation in the fluid}
	
	In the fluid domain, the acoustic pressure ($p$) satisfies the wave equation
	\begin{align}
		\nabla^2 p = \frac{1}{c_0^2}\frac{\partial^2 p}{\partial t^2},
	\end{align}
	where $c_0$ and $\rho_0$ are the intrinsic wave velocity and density of the fluid. Assuming harmonic propagation along the $x$-direction, $p(x,y,t)=\hat p(y)e^{j(\omega t-kx)}$, the governing equation reduces to
	\begin{align}
		\frac{d^2 \hat p}{dy^2} + \left(\frac{\omega^2}{c_0^2}-k^2\right)\hat p = 0,
	\end{align}
	with solution
	\begin{align}
		\hat p(y)=A\cos(k_y y)+B\sin(k_y y),
	\end{align}
	where $k_y^2=\omega^2/c_0^2-k^2$ is the wave number along the $y$-direction.
	Due to symmetry, the acoustic pressure distribution becomes
	\begin{align}\label{eq:acoustic pressure in fluid}
		p=A\cos(k_y y)e^{j(\omega t-kx)}.
	\end{align}

	\textit{ii) Governing equation in the elastic coating}
	
	In the isotropic elastic coating, the displacement field $\mathbf{u}$ satisfies the Navier equation
	\begin{align}\label{eq:elastic wave equation}
		\rho_r \frac{\partial^2 \mathbf{u}}{\partial t^2}
		=
		(\lambda_r + 2\mu_r)\nabla(\nabla\cdot \mathbf{u})
		-
		\mu_r \nabla \times (\nabla \times \mathbf{u}),
	\end{align}
	where $\lambda_r$ and $\mu_r$ are the Lam\'e constants.
	Introducing the Helmholtz decomposition,
	\begin{align}
		\mathbf{u} = \nabla \phi + \nabla \times \mathbf{\Psi},
	\end{align}
	the motion separates into longitudinal component $\phi$ and transverse component $\mathbf{\Psi}$ satisfying
	\begin{align}
		\nabla^2 \phi + k_L^2 \phi = 0, 
		\qquad
		\nabla^2 \mathbf{\Psi} + k_T^2 \mathbf{\Psi} = 0,
	\end{align}
	with the longitudinal ($L$) and transverse ($T$) wave velocities
	\begin{align}
		c_L=\sqrt{\frac{\lambda_r+2\mu_r}{\rho_r}}, 
		\qquad
		c_T=\sqrt{\frac{\mu_r}{\rho_r}}.
	\end{align}
	
	Considering two-dimensional (2D) motion $\mathbf{u}=(u_x,u_y)$ with harmonic dependence $e^{j(\omega t-kx)}$, we introduce the dimensionless variables
	\begin{align}
		X = kx, \qquad Y = \frac{y}{t_r}.
	\end{align}
	Within the long-wavelength and thin-layer limit, i.e., $k t_r\ll 1$, we define the small parameter
	\begin{align}
		\epsilon = k t_r \ll 1.
	\end{align}
	Then the gradient operator becomes
	\begin{align}
		\nabla = \left(k \frac{\partial}{\partial X},\, \frac{1}{t_r}\frac{\partial}{\partial Y}\right)
		= \frac{1}{t_r}\left(\epsilon \frac{\partial}{\partial X},\, \frac{\partial}{\partial Y}\right).
	\end{align}
	The second-order derivatives scale as
	\begin{align}\label{eq:nabla}
		\frac{\partial^2}{\partial x^2} =\frac{\epsilon^2}{t_r^2} \frac{\partial^2}{\partial X^2},
		\frac{\partial^2}{\partial y^2} =\frac{1}{t_r^2} \frac{\partial^2}{\partial Y^2},
		\frac{\partial^2}{\partial x\partial y} = \frac{\epsilon}{t_r^2} \frac{\partial^2}{\partial X\partial Y}.
	\end{align}
	
	Substituting Eq.\ \eqref{eq:nabla} into Eq.\ \eqref{eq:elastic wave equation} yields
	\begin{subequations}
		\begin{align}\label{eq:x-direction}
			\left(\lambda_r+2\mu_r\right)\frac{\epsilon^2}{t_r^2} \frac{\partial^2  u_x}{\partial X^2}
			+\left(\lambda_r+\mu_r\right)\frac{\epsilon}{t_r^2} \frac{\partial^2  u_y}{\partial X\partial Y}&\nonumber\\
			+\mu_r\frac{1}{t_r^2} \frac{\partial^2  u_x}{\partial Y^2}
			&=-\rho_r \omega^2 u_x,\\
			\label{eq:y-direction}
			\left(\lambda_r+2\mu_r\right)\frac{1}{t_r^2} \frac{\partial^2  u_y}{\partial Y^2}
			+\left(\lambda_r+\mu_r\right)\frac{\epsilon}{t_r^2} \frac{\partial^2  u_x}{\partial X\partial Y}&\nonumber\\
			+\mu_r\frac{\epsilon^2}{t_r^2} \frac{\partial^2  u_y}{\partial X^2}
			&=-\rho_r \omega^2 u_y.
		\end{align}
	\end{subequations}
	The above expressions contain six second-order derivative contributions, which scale as
	\begin{align}
		\frac{1}{t_r^2} \partial_{YY}\sim \mathcal{O}(1),~
		\frac{\epsilon}{t_r^2}\partial_{XY}\sim \mathcal{O}(\epsilon),~
		\frac{\epsilon^2}{t_r^2}\partial_{XX}\sim \mathcal{O}(\epsilon^2).
	\end{align}
	Here, the $\mathcal{O}(\cdot)$ notation denotes the scaling order relative to $\epsilon$ after normalizing by the leading-order factor $t_r^{-2}$.
	Accordingly, Eq.\ \eqref{eq:x-direction} and Eq.\ \eqref{eq:y-direction} scale as
	\begin{subequations}
		\begin{align}\label{eq:x-component}
			\text{$x$-component}: \mathcal{O}(1)u_x+\mathcal{O}(\epsilon)u_y+\mathcal{O}(\epsilon^2)u_x,\\
			\text{$y$-component}: \mathcal{O}(1)u_y+\mathcal{O}(\epsilon)u_x+\mathcal{O}(\epsilon^2) u_y.
			\label{eq:y-component}
		\end{align}
	\end{subequations}
	To determine the relative magnitude of the displacement components, we assume a scaling 
	\begin{align}\label{eq:scaling}
		u_x\sim \epsilon^{q}u_y,
	\end{align}
	where $q$ is to be determined. Substituting Eq.\ \eqref{eq:scaling} into Eq.\ \eqref{eq:x-component}, we obtain
	\begin{align}
		\text{$x$-component}: \mathcal{O}(\epsilon^q) u_y	+\mathcal{O}(\epsilon )u_y+\mathcal{O}(\epsilon^{q+2})u_y.
	\end{align}
	Since $q+2>q$, the third term is always higher order than the first and therefore cannot participate in the leading-order balance. The dominant balance must therefore occur between the first two terms, which requires
	\begin{align}
		q=1.
	\end{align}
	Therefore, we have
	\begin{align}\label{eq:order}
		\frac{u_x}{u_y}=\mathcal{O}(\epsilon),
	\end{align}
	indicating that the in-plane and shear-related motions are asymptotically smaller than the normal displacement in the thin-layer limit, so the dominant energy transfer from fluid to solid occurs through normal displacement $u_y$.
	
	{Therefore, we primarily focus on the longitudinal response for the impedance boundary condition.}
	Substituting Eq.\ \eqref{eq:order} into Eq.\ \eqref{eq:y-direction} shows that the leading-order balance is governed solely by the thickness-direction deformation,
	\begin{align}
		\left(\lambda_r+2\mu_r\right)\frac{1}{t_r^2} \frac{\partial^2  u_y}{\partial Y^2}
		=-\rho_r \omega^2 u_y+\mathcal{O}(\epsilon^2),
	\end{align}
	which reduces to the one-dimensional (1D) longitudinal wave equation
	\begin{align}
		\frac{d^2 u_y}{dy^2} + k_L^2 u_y = 0,
	\end{align}
	with general solution
	\begin{align}\label{eq:general solution for solid}
		u_y = A \cos(k_L y) + B \sin(k_L y).
	\end{align}
	
	{We also substitute Eq.\ \eqref{eq:order} into Eq.\ \eqref{eq:x-direction} and obtain
		\begin{align}
			\mu_r\frac{1}{t_r^2} \frac{\partial^2  u_x}{\partial Y^2}
			+\left(\lambda_r+\mu_r\right)\frac{\epsilon}{t_r^2} \frac{\partial^2  u_y}{\partial X\partial Y}
			=-\rho_r \omega^2 u_x+\mathcal{O}(\epsilon^3),
		\end{align}
		which shows that $u_x$ is driven by the dominant longitudinal field $u_y$ and therefore does not represent an independent propagating degree of freedom.
		Instead, $u_x$ remains a higher-order correction of magnitude $\mathcal{O}(\epsilon)u_y$.
	}
	
	The above analysis demonstrates that, within the long-wavelength and thin-layer limit, $k t_r\ll 1$, the full 2D elastodynamic problem asymptotically reduces to a 1D longitudinal response, with in-plane and shear effects entering only at higher order $\mathcal{O}(k t_r)$.

	\textit{iii) Impedance boundary condition at the fluid--solid interface}
	
	We consider harmonic fields with dependence $e^{j(\omega t-kx)}$ in an elastic coating of thickness $t_r$ along the $y$ direction, bounded by a rigid wall at $y=\pm(\ell+t_r)$ and coupled to the fluid at $y=\pm\ell$.
	
	Applying the rigid boundary condition $u_y=0$ at $y=\ell+t_r$, the general solution of the longitudinal displacement in Eq.\ \eqref{eq:general solution for solid} reduces to
	\begin{align}
		u_y(y)=C\sin[k_L(y-\ell-t_r)].
	\end{align}
	The normal stress in the coating is given by
	\begin{align}\label{eq:general stress}
		\sigma_{yy}=(\lambda_r + 2\mu_r)\frac{\partial u_y}{\partial y}
		+
		\lambda_r \frac{\partial u_x}{\partial x}.
	\end{align}
	Using the asymptotic scaling $u_x=\mathcal{O}(\epsilon u_y)$, the second term in Eq.\ \eqref{eq:general stress} is of order $\mathcal{O}(\epsilon^2)$ and can be neglected, yielding
	\begin{align}
		\sigma_{yy} \approx (\lambda_r + 2\mu_r)\frac{d u_y}{d y}.
	\end{align}
	Evaluating the displacement and stress at the fluid--solid interface ($y=\ell$), we obtain
	\begin{align}\label{eq:general expression of displacement and stress}
		\left.u_y\right|_{y=\ell} = -C\sin(k_L t_r),\nonumber\\
		\left.\sigma_{yy}\right|_{y=\ell} = (\lambda_r + 2\mu_r) C k_L \cos(k_L t_r).
	\end{align}
	
	The continuity of normal pressure and velocity at the fluid--solid interface ($y=\ell$) requires
	\begin{align}
		\left.\sigma_{yy}\right|_{y=\ell^+} = \left.-p\right|_{y=\ell^-},
		\qquad
		\left.j\omega u_y\right|_{y=\ell^+} = \left.v_y\right|_{y=\ell^-},
	\end{align}
	which leads to the effective surface impedance at $y=\ell$
	\begin{align}\label{eq:general exp for Z}
		Z_r = \left.\frac{p}{v_y} \right|_{y=\ell^-}= \left.-\frac{\sigma_{yy}}{j\omega u_y}\right|_{y=\ell^+}.
	\end{align}
	Substituting Eq.\ \eqref{eq:general expression of displacement and stress} into Eq.\ \eqref{eq:general exp for Z} yields
	\begin{align}\label{eq:surface impedance}
		Z_r = -j\rho_r c_L \cot(k_L t_r),
	\end{align}
	which provides an asymptotic expression from the full elastodynamic problem to a concise boundary impedance.

	\textit{iv) Asymptotic approximation for the effective wave velocity}
	
	We now derive the effective wave velocity in the water-filled channel lined with the elastic rubber coating.
	Substituting Eq.\ \eqref{eq:acoustic pressure in fluid} into the linearized momentum equation, we obtain the particle vibration velocity along the $y$-direction,
	\begin{align}\label{eq:y velocity}
		v_y& = -\frac{1}{j\omega \rho_0}\frac{\partial p}{\partial y}\nonumber\\
		&= \frac{1}{j\omega \rho_0}k_y A\sin(k_y y)e^{j(\omega t-kx)}.
	\end{align}
	Substituting Eq.\ \eqref{eq:y velocity} into Eq.\ \eqref{eq:general exp for Z}, we obtain
	\begin{align}\label{eq:exact_k_y}
		k_y\tan(k_y\ell)=\frac{j\rho_0 \omega}{Z_r}.
	\end{align}
	Note that Eq.\ \eqref{eq:exact_k_y} is a transcendental equation involving complex numbers,
	for which it is difficult to obtain an explicit analytical solution. In order to provide a clear physical picture, we focus on the propagation of the (0,0)-order wave. 
	
	Within the long-wavelength and thin-layer limit, i.e., $k_y \ell\ll 1$, substituting $\tan(k_y\ell)\approx k_y\ell$ into Eq.\ \eqref{eq:exact_k_y}, we obtain
	\begin{align}
		k_y^2\approx \frac{j\rho_0 \omega}{Z_r\ell}.
	\end{align}
	Using the dispersion $k_0^2=k_x^2+k_y^2$, the wave number along the $x$-direction can be expressed by
	\begin{align}\label{eq:wave number general}
		k_x^2 = k_0^2 \left(1 - \frac{j\rho_0 c_0^2}{\omega Z_r \ell}\right).
	\end{align}
	Substituting Eq.\ \eqref{eq:surface impedance} into Eq.\ \eqref{eq:wave number general}, we obtain
	\begin{align}\label{eq:wave number anal}
		k_x^2 = k_0^2 \left[1+ \frac{\rho_0 c_0^2 \tan(k_L t_r) }{ \omega\rho_r c_L \ell}\right].
	\end{align}
	
	Within the long-wavelength and thin-layer limit, $k_Lt_r\ll 1$, using $\tan(k_L t_r)\approx k_L t_r$, Eq.\ \eqref{eq:wave number anal} can be approximated to
	\begin{align}
		k_x^2 = k_0^2 \left(1 + \frac{t_r}{\ell}\frac{\rho_0 c_0^2}{\rho_r c_L^2}\right).
	\end{align}
	Finally, defining the effective wave velocity as $c_{\mathrm{eff}} = \omega/k_x$, we arrive at
	\begin{align}
		\frac{c_{\mathrm{eff}}}{c_0}\approx
		\left(
		1 + \frac{t_r}{\ell}\frac{\rho_0 c_0^2}{\rho_r c_L^2}
		\right)^{-1/2},
	\end{align}
	which links the effective wave velocity $c_{\mathrm{eff}}$ to the boundary parameters $t_r$, $\rho_r$, $c_L(E_r, \nu_r, \eta_r)$ and channel half-width $\ell$, thereby enabling simultaneous control of phase propagation and energy dissipation.

\begin{acknowledgments}
	The authors thank Prof.\ Kai Zhang, Prof.\ Zhaohe Dai, Prof.\ Yilin Qu, and Dr.\ Chenlei Yu for helpful discussion.
	M.D. acknowledges support from the National Natural Science Foundation of China (Grant No.\ 12602134), the Postdoctoral Fellowship Program of CPSF (Grant No.\ GZB20260485), the Beijing Postdoctoral Research Foundation (Grant No.\ 2026-ZZ-34), and the Fundamental Research Funds for Beijing Municipal Universities (Grant No.\ 312000546325001).
\end{acknowledgments}
\section*{Data Availability}
All study data are included in the article and/or SI Appendix.

\bibliography{apssamp}

\end{document}